# Squeezed Light Generation by Cooper Pair Recombination in a Superconductor-Semiconductor Structure

Sima Buchbinder, Avi Koriat and Alex Hayat

*Faculty of Electrical Engineering, Technion – Israel Institute of Technology, Haifa 32000, Israel*

We theoretically demonstrate squeezed light generation in a compact, electrically driven device utilizing the nonlinear light-matter interaction of Cooper-pair recombination in a hybrid superconductor-semiconductor structure. On the basis of second-order perturbation theory, we developed an effective Hamiltonian which incorporates the electronic component of the interaction as a photon coupling strength coefficient, enabling a model to describe the quadrature squeezing effect of Cooper-pair recombination on the photonic state. We find the squeezing strength to be proportional to the superconducting energy gap squared and device length, as well as increasing with applied voltage. For practical device length, materials and voltage bias, we calculated squeezing values comparable to those of existing squeezed light generation methods when operating below the critical temperature. This theoretically developed concept shows viability for electrically pumped squeezed light generation and can pave the way towards realization of squeezing in compact integrated structures.

## 1. Introduction

Squeezed light describes quantum states for which the uncertainty along one of the quadratures is reduced, compared to classical (coherent) states [1,2]. They are essential in quantum communication, and can been utilized in noise reduction [3], quantum sensing [4], spectroscopy [5], interferometry [6], and gravitational-wave detectors [7,8]. Generation and detection of squeezed light via nonlinear optical processes, such as parametric down conversion [9], optomechanical coupling [10], and optical parametric oscillators [11] have been demonstrated both theoretically and experimentally. However, squeezed states remain challenging to generate [12], and the bulky setups typically needed for the nonlinear optical generation of squeezed light can be an obstacle. Electrically driven squeezed light generation eliminates the need for a strong pump laser and can more readily be used in integrated circuit applications. Superconductor-semiconductor devices are a rapidly evolving field, combining the technologically mature field of semiconductors with the exotic properties of superconductors, which are utilized in quantum computing, information and communication. The electrically driven Cooper pair recombination in superconducting light emitting diodes (SLEDs) has been experimentally observed [13], and theoretically shown to produce polarization-entangled photon pairs [14,15]. This superconductor-enhanced light emission can also be utilized in two-photon amplification in waveguides [16] and Bell-state analyzers [17].

Here, we show theoretically that the nonlinear optical interaction of Cooper-pair recombination in a hybrid superconductor-semiconductor waveguide can be utilized to generate squeezed light, when seeded by a coherent state. This presents a method for the electrically driven generation of squeezing, which is crucial for the integration of optoelectronic circuits for quantum communication. We developed a model to describe the effective Hamiltonian of the nonlinear interaction effect on the photons, and show that the squeezing is proportional to the superconducting energy gap squared $\Delta^2$, such that the state becomes more squeezed as the temperature decreases, and above the critical temperature the squeezing vanishes. The squeezing increases with increasing waveguide length and applied voltage. In addition, the photon energy spectrum of the squeezed light is limited to the upper and lower bounds of the energies for which Cooper pair recombination can take place, determined by the hole and Cooper pair energies in the squeezing spectrum, as well as the various the Fermi-Dirac distribution functions.

## 2. Theoretical Model

Given a superconductor-semiconductor structure (Fig. 1a), we investigate the propagation of a laser pulse such that two photons are emitted via Cooper pair recombination (Fig. 1b). The Cooper-pair recombination is described by the light-matter interaction Hamiltonian given by [18]:

$$H_I = \sum_{k,q,\sigma} B_{k,q} b_{k-q,\sigma} c_{k,\sigma} a_q^\dagger + B_{k,q}^* b_{k-q,\sigma}^\dagger c_{k,\sigma}^\dagger a_q \,, \tag{1}$$

where $b_{k,\sigma}(c_{k,\sigma})$ is the annihilation operator of a hole (electron) with a wave number $k$, spin $\sigma$, and excitation energy $\xi_k^p$ ($\xi_k^n$), and $a_q$ is the annihilation operator of a photon with a wave number $q$ and energy $E_{ph}$.

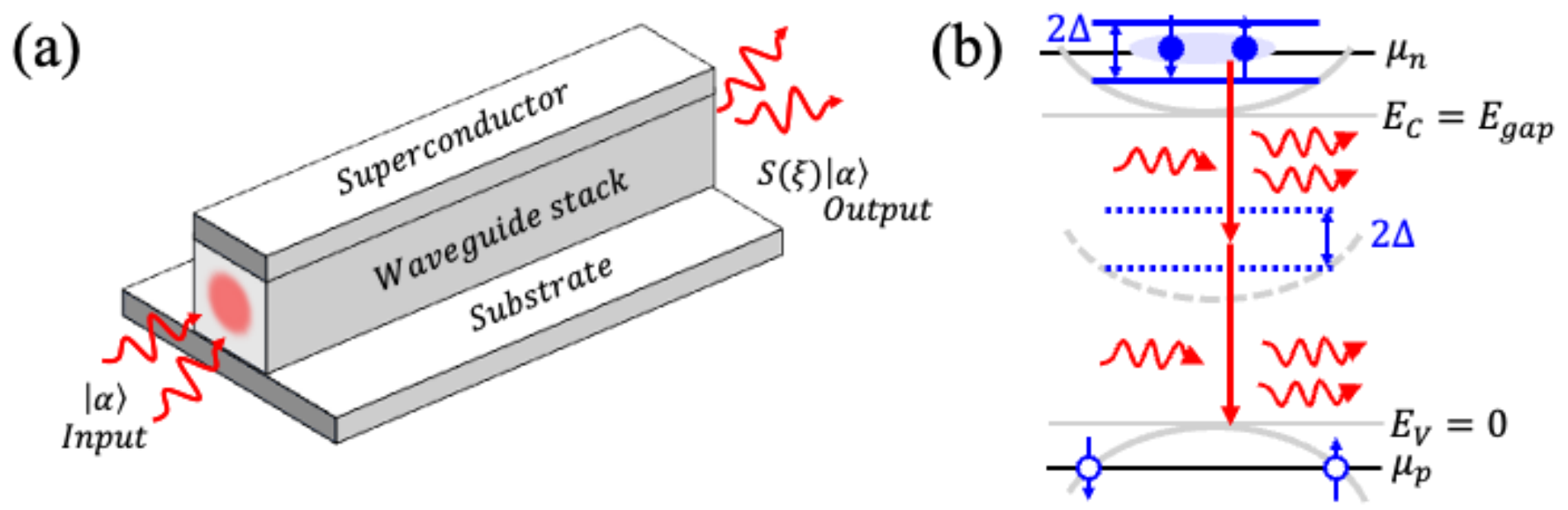

Fig. 1: (a) Schematic of the hybrid waveguide device. (b) Cooper pair recombination. The electron Cooper pair and pair of holes recombine via a virtual level, such that the total energy is conserved. $\mu_{n(p)}$ is the electron (hole) quasi-Fermi level.

The energy bands of the system are shown schematically in Fig. 1b. The top of the valence band $E_V$ is the system's reference level, and the bottom of the conduction band is $E_C = E_{gap}$, where $E_{gap}$ is the semiconductor's energy gap. The distance between the quasi-Fermi levels is equal to the applied voltage, $V_a$, which is assumed to be larger than $E_{gap}$. It is also assumed that both quasi-Fermi levels move evenly with applied voltage, such that $\mu_n = E_C + \frac{eV_a - E_{gap}}{2}$ and $\mu_p = E_V - \frac{eV_a - E_{gap}}{2}$. The electron and hole energies are given by the known parabolic dispersion relation: $E_k^n = E_C + \frac{\hbar^2 k^2}{2m_e} = E_{gap} + \frac{\hbar^2 k^2}{2m_e}$, $E_k^p = E_V - \frac{\hbar^2 k^2}{2m_h} = -\frac{\hbar^2 k^2}{2m_h}$. The excitation energies are related to the energy levels by: $\xi_k^p = E_k^p - \mu_p$ , $\xi_k^n = E_k^n - \mu_n$ .

The process of Cooper pair recombination in the interaction picture is described by the second-order time evolution operator, given by [19]:

$$U^{(2)}(t_0 + \Delta t, t_0) = \frac{1}{i\hbar}\int_{t_0}^{t} dt_2 \frac{1}{i\hbar}\int_{t_0}^{t_2} dt_1 \bar{H}_I(t_2)\bar{H}_I(t_1) \ , \quad (2)$$

where $\bar{H}_I(t) = e^{iH_0(t-t_0)/\hbar} H_I e^{-iH_0(t-t_0)/\hbar}$, $H_I$ is as defined in Eq. (1), and $H_0$ is the non-interactive Hamiltonian as given in [18].

Under the assumption of a short interaction time $\Delta t$, the integrand of the 2nd order time evolution operator in Eq. (2) is taken as the effective Hamiltonian in the photon space, sampled at time $t_2$, because the Cooper pair recombination occurs via a virtual state [14], with a very short interaction time. In addition, we express the electronic part of the interaction as a constant which describes its effect on the photons: Similarly to the development of the nonlinear optical susceptibility [20], we incorporate the matrix element of Cooper pair recombination into the effective Hamiltonian. Thus, an approximation of the interaction sampled at a time $t_2$ of our choosing is obtained:

$$H_{eff}^{ph} = \langle \chi_f |_{eh} \left[ \frac{1}{i\hbar}\int_{-\infty}^{t_2} dt_1 \bar{H}(t_2)\bar{H}(t_1) \right]_{t_2=0} |\chi_i\rangle_{eh} + \mathrm{H.C.}. \quad (3)$$

This can be likened to the parametric approximation commonly used in quantum optics [20,21], in which an intense, undepleted and coherent optical pump field is approximated as classical, and as such the pump's operator in the system Hamiltonian is replaced with a number. For the approximation in Eq. (3) to hold, the population change due to radiative recombination is assumed to be negligible compared to the steady state concentration of holes and Cooper pairs. This requires that the applied current must be high enough to enable a large number of Cooper pairs in the CB and holes in the VB available for recombination, while not exceeding the critical current. Additionally, the device must operate below the critical current and temperature for the approximation to be valid.

The initial electronic state $|\chi_i\rangle_{eh} = |BCS\rangle_e|FS\rangle_h$ describes a BCS superconducting condensate for the electrons and a Fermi sea of holes. The final state, denoted by $|\chi_f\rangle = |BCS-2\rangle_e|FS-2\rangle_h$ is a state for which a pair of photons has been created by the recombination of a Cooper pair with a pair of holes. The full state of the system is of the form $|\chi\rangle_{ph}|\chi\rangle_{eh}$, such that its evolution $U^{(2)}|\chi_i\rangle_{ph}|\chi_i\rangle_{eh}$ becomes $U_{eff}|\chi_i\rangle_{ph}$ under our approximation, where $U_{eff}$ is the time evolution operator developed from $H_{eff}^{ph}$. The sampling time was taken to be $t_2 = 0$. The effective Hamiltonian obtained from Eq. (3) using similar assumptions to previous works [18] is (full development can be found in the provided supplementary material [22]):

$$H_{eff}^{ph} = \sum_{q_1}\sum_{q_2} \beta_{q_1} a_{q_1}^{\dagger} a_{q_2}^{\dagger} + \beta_{q_1}^{*} a_{q_1} a_{q_2} \,, \tag{4}$$

where the nonlinear interaction coefficient is:

$$\beta_q = \frac{2Al}{(2\pi)^2}\int_0^{k_{max}} \left\{ k^2 \frac{B^2\Delta^2}{\left(\xi_k^{QP}\right)^2} \cdot f_{FD}\left(\xi_k^{p}\right)^2 \times \right.$$

$$\left. \times \left[ \frac{2f_{FD}\left(\xi_k^{QP}\right)^2 - 3f_{FD}\left(\xi_k^{QP}\right) + 1}{-\xi_k^{p} + eV_a - E_{ph} + \xi_k^{QP}} + \frac{2f_{FD}\left(\xi_k^{QP}\right)^2 - f_{FD}\left(\xi_k^{QP}\right)}{-\xi_k^{p} + eV_a - E_{ph} - \xi_k^{QP}} \right] \right\} dk, \tag{5}$$

$\xi_k^{QP} = \sqrt{(\xi_k^{n})^2 + \Delta^2}$ is the excitation energy of the BCS quasiparticle relative to $\mu_n$, $f_{FD}(E)$ is the Fermi-Dirac distribution function, $A$ is the active area cross-section of the waveguide, and $l$ is a small segment of the waveguide such that the propagation time through the section is sufficiently short to satisfy the requirement on $\Delta t$.

With a practical scattering induced broadening term $\eta$ introduced in the denominators [18], Eq. (5) becomes:

$$\beta_q = \frac{2Al}{(2\pi)^2}\left|\int_0^{k_{max}}\left\{k^2\frac{B^2\Delta^2}{\left(\xi_k^{QP}\right)^2 + i\eta}\cdot f_{FD}\left(\xi_k^p\right)^2 \times \right.\right.$$

$$\left.\left.\times\left[\frac{2f_{FD}\left(\xi_k^{QP}\right)^2 - 3f_{FD}\left(\xi_k^{QP}\right) + 1}{-\xi_k^p + eV_a - E_{ph} + \xi_k^{QP} + i\eta} + \frac{2f_{FD}\left(\xi_k^{QP}\right)^2 - f_{FD}\left(\xi_k^{QP}\right)}{-\xi_k^p + eV_a - E_{ph} - \xi_k^{QP} + i\eta}\right]\right\}dk\right|. \tag{6}$$

Observing the effective Hamiltonian in Eq. (4), we note that taking its full time evolution would yield the multi-mode squeezing operator, as defined in [23]. For simplicity, we shall consider the case of a single mode $q$ seed, which can be filtered spectrally. The effective time-independent Hamiltonian in Eq. (4) becomes $H_{eff}^{ph} = \beta_q\left(a_q^\dagger\right)^2 + \beta_q^* a_q^2$, which is similar in its form to that of degenerate parametric amplification [21]. Therefore, the time evolution operator is:

$$U_{eff}(t) = e^{-i\frac{H_{eff}^{ph}}{\hbar}t} = e^{-\frac{i}{\hbar}\left[\beta_q\left(a_q^\dagger\right)^2 + \beta_q^* a_q^2\right]t}, \tag{7}$$

where $t$ is the total propagation time in the waveguide. This can be expressed as $U_{eff}(t) = S\left(\frac{2i\beta_q}{\hbar}t\right)$, where $S(\xi)$ is the squeezing operator as defined in [1]. The time evolution operator in Eq. (7) and an initial photon state $|\chi_0\rangle_{ph} = |\alpha_q\rangle$ result in the rotated quadrature uncertainty $\left(\Delta Y_{1,2}\right)_t^2 = \langle Y_{1,2}^2\rangle_t - \langle Y_{1,2}\rangle_t^2 = e^{\mp\frac{4t|\beta_q|}{\hbar}}$, where $\langle\hat{O}\rangle_t = \langle\chi_0|_{ph}U_{eff}^\dagger(t)\hat{O}U_{eff}(t)|\chi_0\rangle_{ph}$, and the rotated quadrature operators are

$$Y_1(\theta) = ae^{-\frac{i\theta}{2}} + a^\dagger e^{\frac{i\theta}{2}}$$
$$Y_2(\theta) = i\left(a^\dagger e^{\frac{i\theta}{2}} - ae^{-\frac{i\theta}{2}}\right), \tag{8}$$

with a rotation angle of $\theta = \arg\left(\frac{2i\beta_q}{\hbar}t\right) = \frac{\pi}{2}$.

For clarity, the squeezing is presented in dB units [21,24,25]:

$$\text{Squeezing(dB)} \equiv -10\log(\Delta Y_1^2)_t = \frac{10}{\ln 10}\cdot\frac{4t|\beta_q|}{\hbar}. \tag{9}$$

## 3. Discussion

For realistic device parameters, the presented framework exhibits squeezing of several dB for an ideal waveguide with a length of up to 10s of mm. This is comparable to results obtained by conventional methods of squeezed light generation [9,10,26] with similar typical lengths. We note that in the current framework, in the absence of a seed pulse there will be spontaneous emission of energy-entangled photon pairs [27–30]. However, entanglement based on evolution of a coherent seed is unfeasible due to spontaneous emission in the device.

As a specific example, we studied the case of a hybrid waveguide with a GaAs semiconductor structure (for which $E_{gap} = 1.42eV$) and a Nb superconducting contact. The temperature dependence of $E_{gap}$ was neglected, and the critical temperature of the superconductor was taken to be $T_c = 9.25K$. The active area cross section of the modeled device is $A = 1\mu\text{m}\times 500\text{nm}$, and unless stated otherwise device length in the numerical calculations is $L = 1mm$. By default, the relaxation parameter was taken to be $\eta \approx 0.4meV$, assuming a typical relaxation time of $\tau \approx 0.1ps$ [31,32]. The superconducting energy gap $\Delta(T)$ was taken empirically [33] to be $\Delta(T) = \Delta_0 \tanh\left(1.74\sqrt{1-\frac{T}{T_c}}\right)$. The superconducting gap dependence on the applied current is neglected, since for the waveguide typical geometry the critical current is estimated to be in the order of 2A – much higher than typical experimental values.

Since the time evolution of the effective Hamiltonian yields a squeezed coherent state, the Wigner function is plotted using the formula presented in [34] :

$$W_{|\alpha_0,\xi_0\rangle} = \frac{2}{\pi}\exp\left(-\frac{\gamma_x^2}{2\sigma_s^2} - \frac{\gamma_y^2}{2\sigma_a^2}\right) . \tag{10}$$

This formula assumes a general squeezed coherent state $|\alpha_0, \xi_0\rangle$ defined by two complex numbers: the displacement $\alpha_0$ and the squeezing parameter $\xi_0$. The squeezing parameter is expressed using its phase and amplitude $\xi_0 = r_0 e^{i\theta_0}$. The Wigner function above also utilizes the notations:

$$\gamma_x = (X_1 - \text{Re}\{\alpha_0\})\cos\left(\frac{\theta_0}{2}\right) + (X_2 - \text{Im}\{\alpha_0\})\sin\left(\frac{\theta_0}{2}\right) , \tag{10.1}$$

$$\gamma_y = -(X_1 - \mathrm{Re}\{\alpha_0\}) \sin\left(\frac{\theta_0}{2}\right) + (X_2 - \mathrm{Im}\{\alpha_0\}) \cos\left(\frac{\theta_0}{2}\right) , \tag{10.2}$$

$$\sigma_s = \frac{e^{-r_0}}{2} , \tag{10.3}$$

$$\sigma_a = \frac{e^{r_0}}{2} , \tag{10.4}$$

where the quadrature operators $X_{1,2}$ are given by Eq. (8): $X_{1,2} = Y_{1,2}(\theta = 0)$.

The proposed scheme produces quadrature squeezing such that the uncertainty in the rotated quadrature $Y_1$ is reduced while the $Y_2$ uncertainty is increased (Fig. 2). As the temperature is increased or the device length decreases, the state becomes less squeezed, and its Wigner function approaches that of a coherent state. The squeezing reduces as temperature increases because the nonlinear interaction parameter $\beta_q$ is proportional to $\Delta^2$, which in turn decreases with higher temperatures. As for the effect of the device length, Eq. (9) shows that the squeezing is proportional to the propagation time in the waveguide. For a given device of length $L$, this interaction time is around $t = \frac{L}{v}$, where $v$ is the group velocity of an electromagnetic signal in the waveguide. Thus, as the device length increases, the state will become more squeezed.

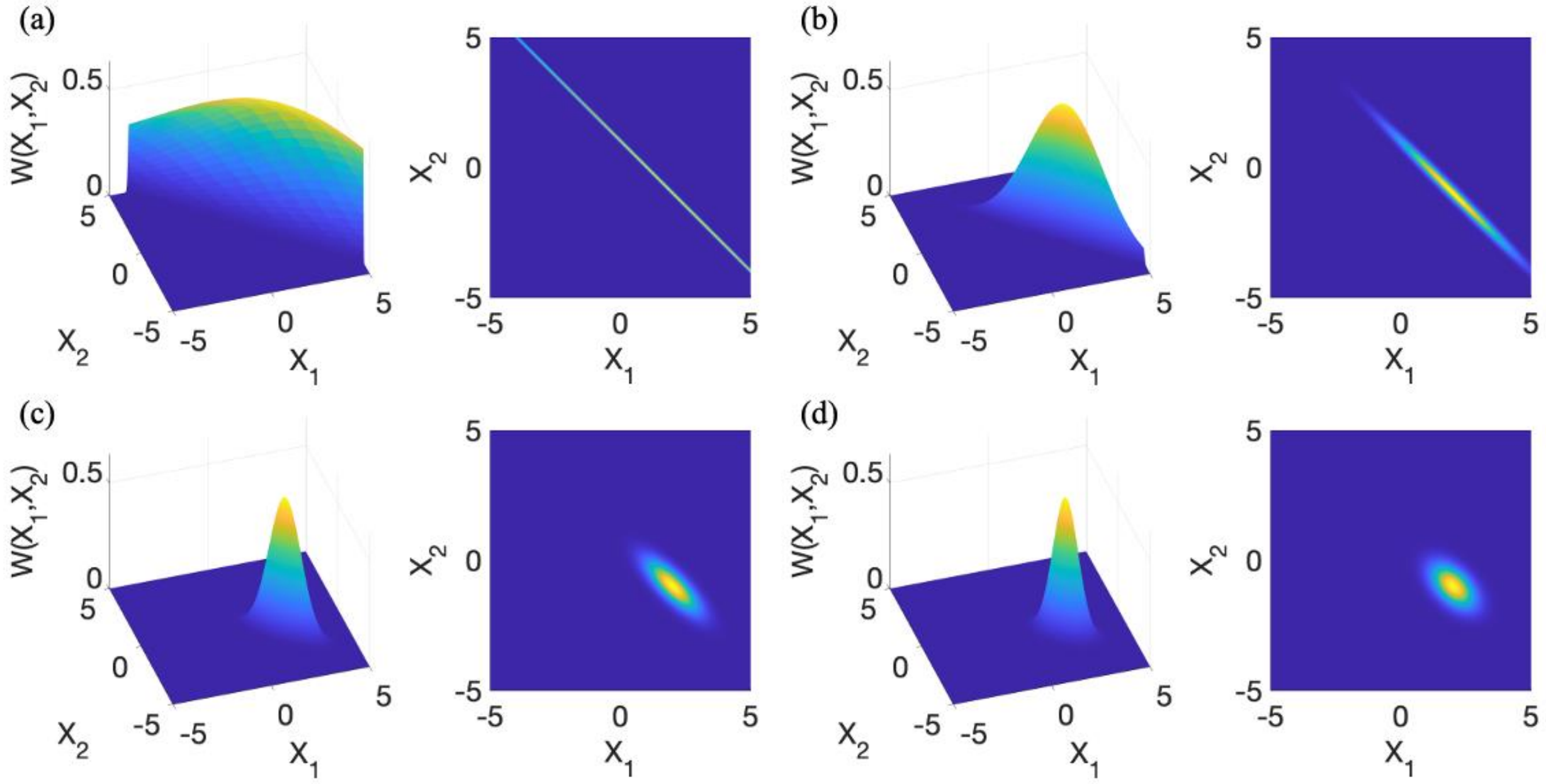

Fig. 2: Wigner functions of the state in the quadrature $(X_1, X_2)$ space from an angle (left) and from above (right) for (a) $T = 0.25T_C, L = 5mm$, (b) $T = 0.75T_C, L = 5mm$, (c) $T = 0.25T_C, L = 1mm$, (d) $T = 0.75T_C, L = 1mm$ , assuming an initial state $|\alpha_q\rangle = |2 - i\rangle$, $V_a - E_{gap} = 7mV$ and $E_{ph} = 1.425eV$

This scheme cannot be tuned such that the uncertainty is reduced for a specific axis. Similarly to stimulated emission in a diode laser [35], the phase between the seed and the output is fixed because there is no pump field, but rather an injected carrier concentration. As such, the nonlinear interaction parameter has no tunable phase dependence, and the rotation angle of the quadratures $\theta$ for which the uncertainty is reduced is constant.

The temperature dependence of the squeezing as seen in Fig. 2 is consistent throughout all photon energies (Fig. 3), and as expected by the dependence on $\Delta$, above the critical temperature the squeezing vanishes (Fig. 3a). The squeezing also increases with applied voltage throughout all the photon energy spectrum – as the voltage is increased, more current is injected into the junction, causing more recombination of Cooper pairs, such that the nonlinear coefficient $\beta_q$ increases.

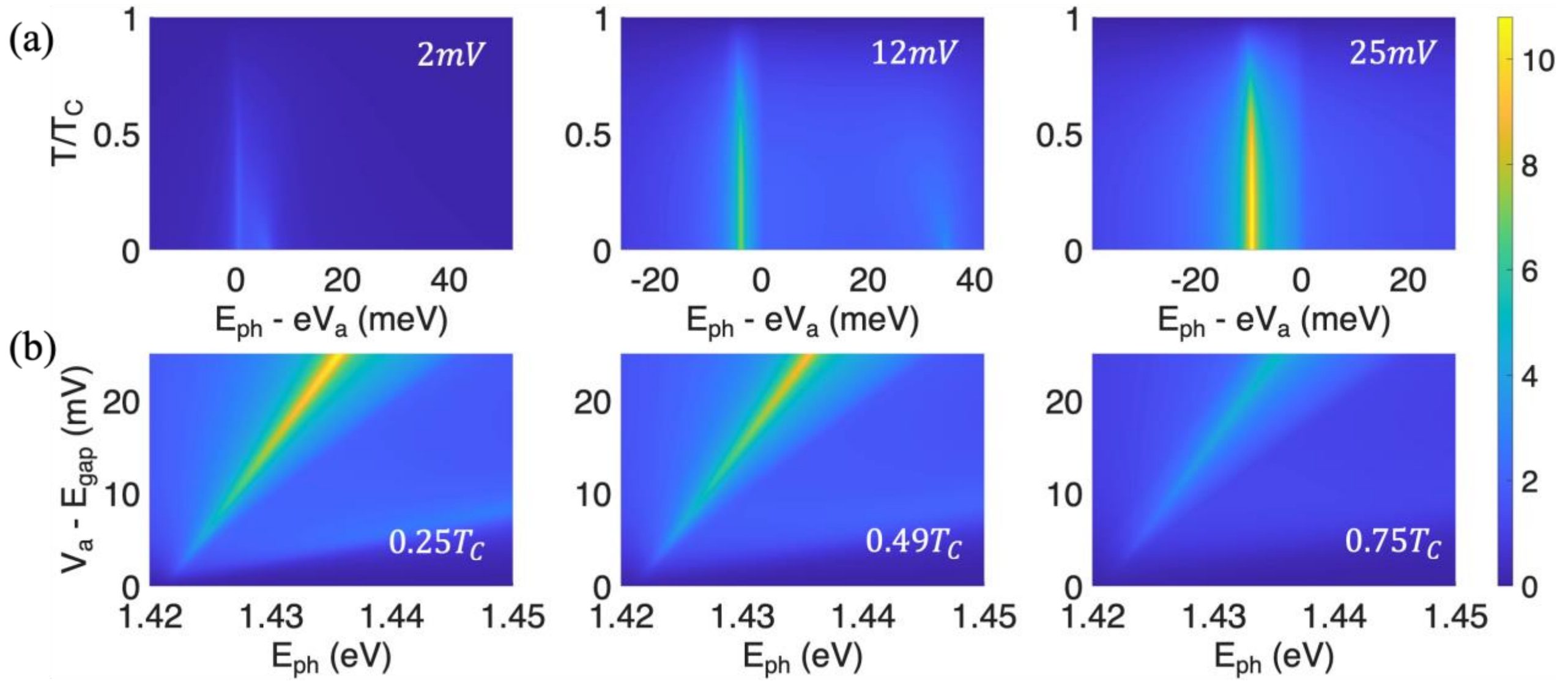

Fig. 3: (a) Squeezing in dB vs. temperature and photon energy for several applied voltages $V_a - E_{gap}$. (b) Squeezing in dB vs. applied voltage and photon energy for several temperatures $T$.

The spectrum of the squeezed light presents a peak at $E_{ph} = eV_a + \Delta - \xi^p_{k_{fn}}$, for which there is a resonance of the quasiparticles in the intermediate state of 2nd order perturbation (Fig. 6). For photon energies larger than $eV_a$ (Fig. 3a) the squeezing falls significantly, as the overall emission reduces. A second (local) maximum of the squeezing is present at $E_{ph} = eV + \xi^n_{k_{fp}}$ due to the intermediate state meeting a maximum in the Fermi-Dirac multiplication product in Eq. (6), however this peak broadens substantially with increasing temperature (Fig. 3b) due to the broadening of the Fermi-Dirac distribution. For energies higher than $E_{ph} = eV_a + \xi^n_{k_{fp}}$ or lower than $E_{ph} = eV_a + \Delta - \xi^p_{k_{fn}}$, the squeezing approaches zero. The location of these bounds of the squeezing increases linearly with the applied voltage (Fig. 3b), due to the linear dependence between the quasi-Fermi-levels (present in $\xi_k^{n,p}$) and $V_a$.

When observing the effect of the device length on the squeezing spectrum, we find that the strength of the squeezing increases with increasing device length (Fig. 4), however when the spectrum of every device length is normalized to its maximal value, we find that the energy dependence of the squeezing does not change with device length.

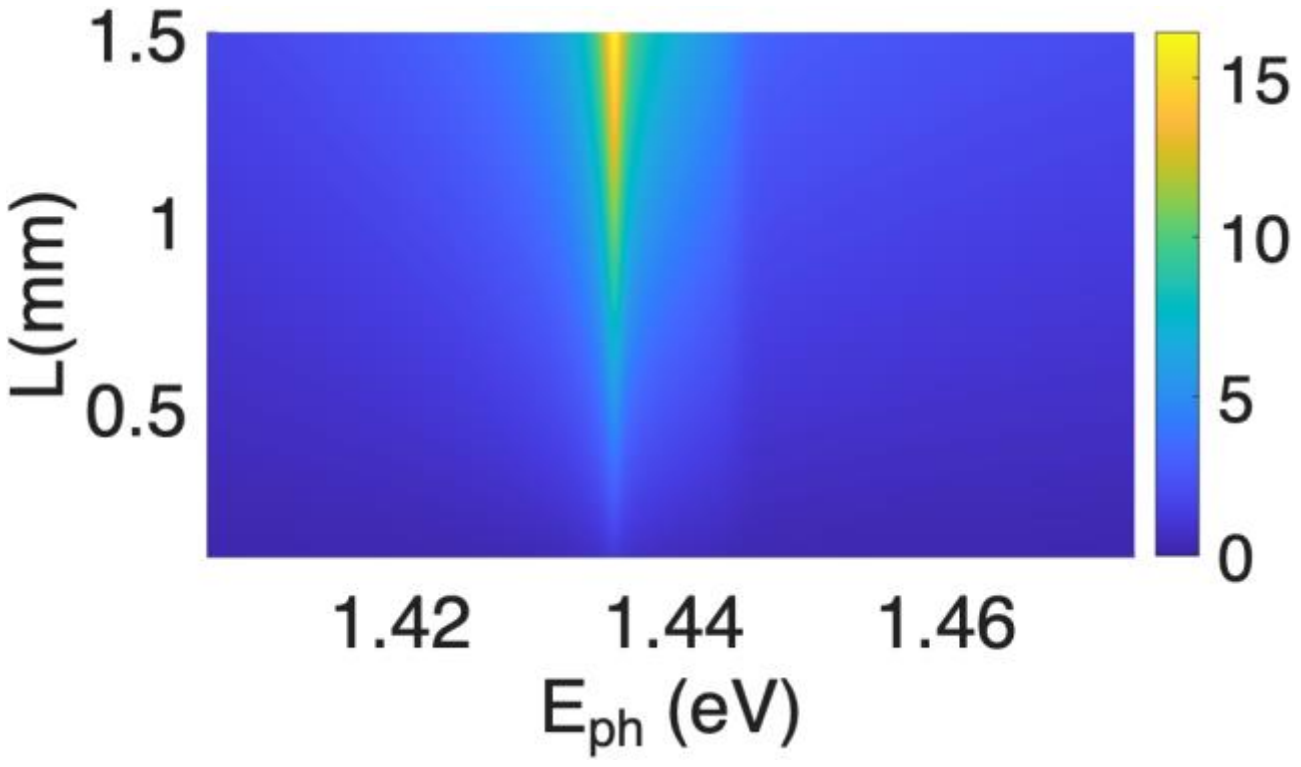

Fig. 4: Squeezing in dB vs. device length and photon energy for applied voltage $eV_a - E_{gap} = 25meV$ and temperature $T = 0.25T_C$

We find that a larger broadening $\eta$ leads to a wider spectral signature of the squeezing but lowers the maximum amplitude of the squeezing, as expected (Fig. 5).

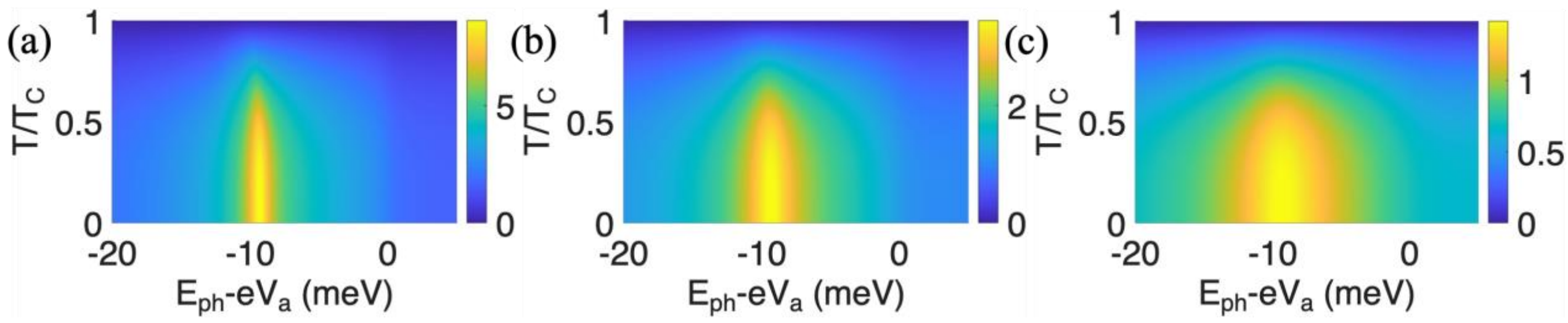

Fig. 5: Squeezing in dB vs. temperature and photon energy for applied voltage $eV_a - E_{gap} = 25meV$ for (a) η=0.5meV, (b) η=1meV and (c) η=2meV

Contrary to what one might expect, the spectral bounds in Figs. 3-5 do not fully correspond to resonant energies in the denominators of the nonlinear interaction coefficient seen in Eq. (6). This can be explained by examination of the k-spectral behavior of $\left|\frac{d\beta_q}{dk}\right|$, which is given by:

$$\left|\frac{d\beta_q}{dk}\right| = \frac{2Al}{(2\pi)^2}\left|k^2 \frac{B^2\Delta^2}{\left(\xi_k^{QP}\right)^2 + i\eta} \cdot f_{FD}\left(\xi_k^p\right)^2 \times \right.$$
$$\left. \times \left[\frac{2f_{FD}\left(\xi_k^{QP}\right)^2 - 3f_{FD}\left(\xi_k^{QP}\right) + 1}{-\xi_k^p + eV_a - E_{ph} + \xi_k^{QP} + i\eta} + \frac{2f_{FD}\left(\xi_k^{QP}\right)^2 - f_{FD}\left(\xi_k^{QP}\right)}{-\xi_k^p + eV_a - E_{ph} - \xi_k^{QP} + i\eta}\right]\right|, \tag{11}$$

and its resonant points.

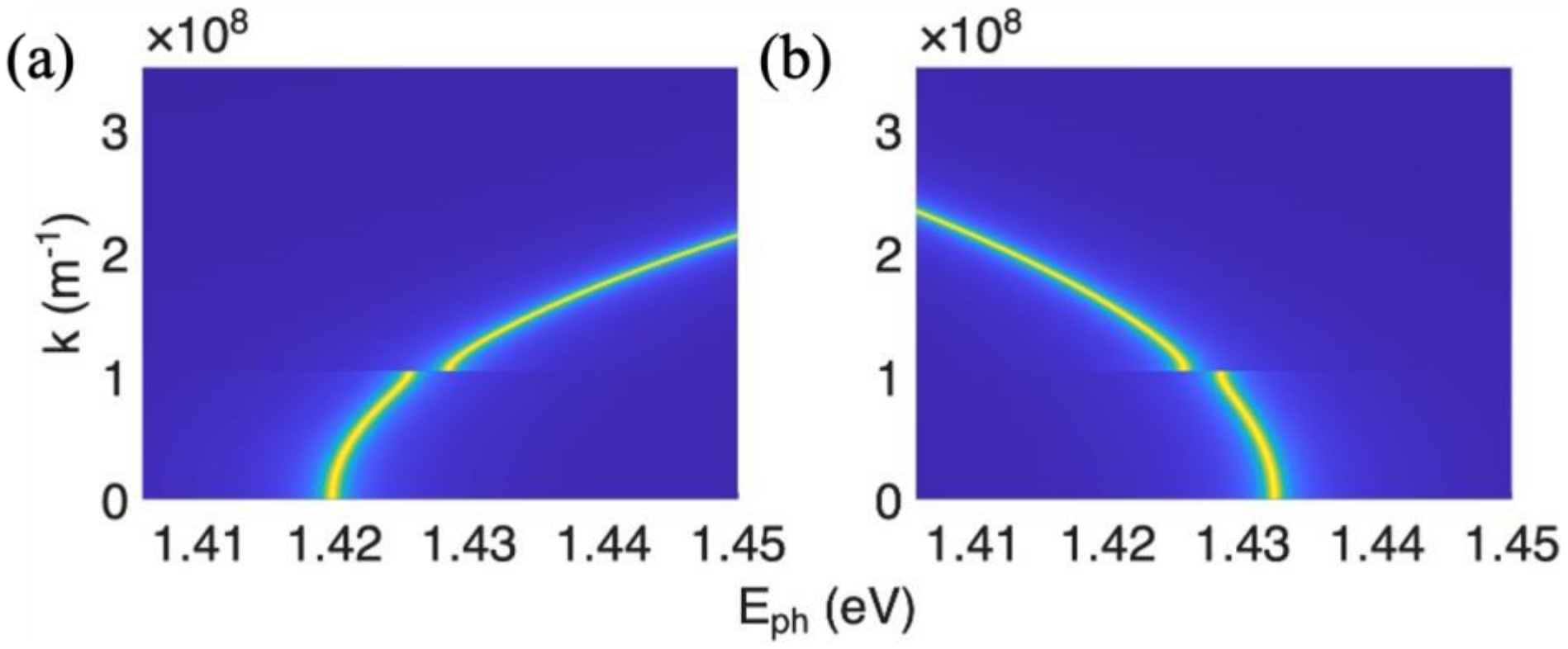

Fig. 6: Resonance points of Eq. (11): (a) positive quasiparticle energy $\left(-\xi_k^p + eV_a - E_{ph} + \xi_k^{QP} + i\eta\right)^{-1}$ and (b) negative quasiparticle energy $\left(-\xi_k^p + eV_a - E_{ph} - \xi_k^{QP} + i\eta\right)^{-1}$ plotted vs. photon energy and wave vector, for constant applied voltage $V_a = E_{gap} + 12mV$ and temperature $T = 0.25T_C$

We observe that while one resonance spans energies from $E_{gap}$ and higher (Fig. 6a), and the other spans energies from $eV_a$ and lower (Fig. 6b). Furthermore, the resonance point of both denominators features an energy gap at the Fermi wave vector $k_{f_n}$, due to the superconducting energy gap affecting $\xi_k^{QP}$. This gap can be shown to span from roughly $E_{ph} = \mu_n$ to $E_{ph} = eV_a + \Delta - \xi_{k_{f_n}}^p$. Once the resonances are multiplied by the Fermi-Dirac distributions, only the electron-like quasiparticle ($k > k_{f_n}$) branch of Fig. 6a and the hole-like quasi particle ($k < k_{f_n}$) branch of Fig. 6b remain, such that the superconducting gap in the spectrum of $\left|\frac{d\beta_q}{dk}\right|$ disappears (Fig. 7).

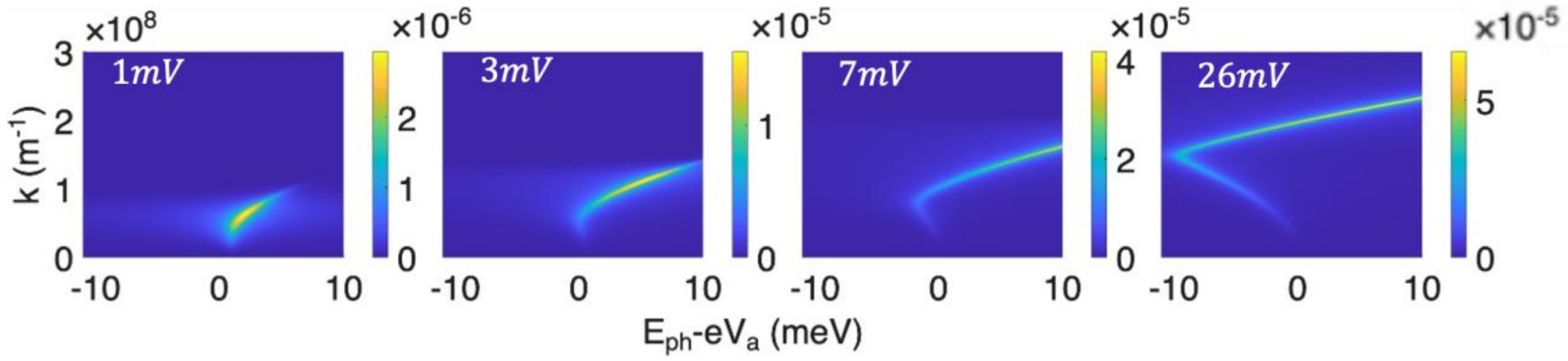

Fig. 7: $\left|\frac{d\beta_q}{dk}\right|$vs. photon energy $E_{ph}$ and wave vector $k$, for constant temperature $T = 0.25T_C$ and several applied voltages $V_a - E_{gap}$.

Therefore, the superconducting energy gap does not appear as a feature in the squeezing spectrum after integration over k (Fig. 7). Note the increase in the value of $\left|\frac{d\beta_q}{dk}\right|$ with applied voltage – as previously stated, the nonlinear interaction factor increases with voltage due to increased recombination. The left hand "knee" of the $\left|\frac{d\beta_q}{dk}\right|$ spectrum, where the electron-like and hole-like quasiparticle branches meet, corresponds to the photon energy $E_{ph} = eV_a + \Delta - \xi^p_{k_{fn}}$ (global maximum and lower bound of squeezing spectrum as seen in Figs. 3-5). The point where $\left|\frac{d\beta_q}{dk}\right|$ begins to reduce with photon energy corresponds to $E_{ph} = eV_a + \xi^n_{k_{fp}}$ (upper bound of squeezing spectrum).

There are several practical requirements on the proposed scheme which must be taken into consideration when discussing experimental feasibility. Firstly, in this manuscript we have discussed only the two-photon (2P) emission in the device resulting from Cooper Pair recombination. However, in these devices unwanted one-photon (1P) emission is present as well. In order to observe significant squeezing, the uncorrelated 1P emission must be reduced compared to the 2P emission. There are spectral differences between the 1P and 2P emission, and therefore filtering can be utilized in the waveguide's design in order to increase the signal-to-noise ratio when measuring the squeezing in an experimental setup. This topic has been discussed in previous works [13,36,37] in the context of enhancing 2P emission, which propose implementing bandgap engineering via quantum well or superlattice layers in the waveguide semiconductor stack in order to enhance the Cooper pair injection. Another option in the waveguide's design in order to limit 1P emission is optical filtering via a cavity.

Another consideration in the waveguide design is the overlap between the waveguide transverse mode, the PN junction active region (where radiative recombination occurs) and the Cooper Pair penetration depth. Since the Cooper Pair recombination takes place in the overlap area, the efficiency of squeezed light generation will increase with the overlap. In previous work relating to two-photon amplification in a similar structure [38], the utilization of plasmonic waveguides is suggested as a way to increase the overlap, but due to large losses this solution would be inappropriate for the case of squeezed light generation.

Next, we consider the material factors of dispersion and propagation loss. Dispersion in the waveguide, will cause a broadening of the input pulse. This will effectively increase the interaction time but lower the input intensity. Since the Cooper pair recombination process has a linear dependence on the interaction time and a quadratic dependence on the intensity [16], overall the efficiency of the process will decrease, causing an effective reduction of the nonlinear interaction parameter $\beta_q$. Assuming a single mode waveguide, for the device's working temperatures [39] there will be a broadening of approximately 20% for a 100fs input pulse propagating in a 1mm waveguide [40], and the efficiency is expected to reduce by a similar factor.

Variations in the interface quality will introduce propagation losses. In this manuscript we have calculated the squeezing for an ideal, lossless waveguide, but losses can greatly affect the experimental generation of squeezed light [41]. While the lossless squeezing, as presented in Eq. (9) of the manuscript is given by $S_{ideal} = -10\log\left(e^{-\frac{4t\beta}{\hbar}}\right)$, the squeezing in the presence of losses $S_{practical}$ is [42,43]:

$$S_{practical} = -10\log\left(1 - T + Te^{-\frac{4t\beta}{\hbar}}\right), \tag{12}$$

here $T$ is the total transmission efficiency. Given a propagation loss of $\alpha\left[\frac{dB}{cm}\right]$, this efficiency is $T = 10^{-\alpha L/10}$. For typical values, we can expect a reduction of several dB for a mm-scale device. Since the interaction time increases with $L$ as seen in Eq. (9), while the efficiency decreases, we find that for a practical device the squeezing will increase with waveguide length only up to a certain point, from which the losses become dominant.

## 4. Conclusions

In conclusion, we have shown that the hybrid semiconductor-superconductor waveguide can be used as an electrically-driven source of squeezed coherent light via the nonlinear interaction of Cooper-pair recombination. We presented a theoretical model for the squeezed light generation by developing an effective Hamiltonian describing the interaction in the photon space, which incorporates electronic component of the interaction as a photon coupling strength coefficient. In this model, we showed a close relation between the squeezing intensity and the superconducting energy gap and observed the existence of resonance photon energies for which the squeezing will be strongest for a given temperature and applied voltage. We showed that for a coherent input, the output state becomes more squeezed as the temperature decreases and device length and applied voltage increase, as shown by both the Wigner functions and the spectral distribution of the squeezed light. We demonstrated the changes of the squeezing spectrum with temperature and applied voltage, such as a shift of the squeezed light's spectrum towards higher photon energy with increased voltage, but not with increased temperature. We calculated squeezing values comparable to existing methods for realistic device parameters in a lossless waveguide, and discussed the practical feasibility of the proposed scheme. This new squeezed light source concept can enable future integrated photonic devices for quantum technologies such as metrology and communication.